\catcode`\@=11

\newif\if@fewtab\@fewtabtrue

{\count255=\time\divide\count255 by 60
\xdef\hourmin{\number\count255}
\multiply\count255 by-60\advance\count255 by\time
\xdef\hourmin{\hourmin:\ifnum\count255<10 0\fi\the\count255}}
\def\ps@draft{\let\@mkboth\@gobbletwo
    \def\@oddhead{}
    \def\@oddfoot
       {\hbox to 7 cm{$\scriptstyle Draft\ version:\ \draftdate$
       \hfil}\hskip -7cm\hfil\rm\thepage \hfil}
    \def\@evenhead{}\let\@evenfoot\@oddfoot}

\def\ceqno{\global\@fewtabfalse
    \ifcase\@eqcnt \def\@tempa{& & &}\or \def\@tempa{& &}
      \or \def\@tempa{&}
      \or\def\@tempa{}\fi\@tempa
{\rm(\theequation)}}

\def\aeqno#1{\global\@fewtabfalse
    \ifcase\@eqcnt \def\@tempa{& & &}\or \def\@tempa{& &}
      \or \def\@tempa{&}
      \or\def\@tempa{}\fi\@tempa
{\rm(\theequation,#1)}}

\def\label#1{\ifnum\draftcontrol=1
 \global\def\draftnote{$\scriptstyle #1$}\fi
 \@bsphack\if@filesw {\let\thepage\relax
   \def\protect{\noexpand\noexpand\noexpand}%
\xdef\@gtempa{\write\@auxout{\string
      \newlabel{#1}{{\@currentlabel}{\thepage}}}}}\@gtempa
   \if@nobreak \ifvmode\nobreak\fi\fi\fi
  \@esphack}

\def\alabel#1#2{\label{#1}\global\@fewtabfalse
    \ifcase\@eqcnt \def\@tempa{& & &}\or \def\@tempa{& &}
      \or \def\@tempa{&}
      \or\def\@tempa{}\fi\@tempa
{\hbox to 3cm{\phantom{\rm(\theequation,#2)}
\draftnote \hfil}\hskip -3cm {\rm(\theequation,#2)}}}

\def\clabel#1{\label{#1}\global\@fewtabfalse
    \ifcase\@eqcnt \def\@tempa{& & &}\or \def\@tempa{& &}
      \or \def\@tempa{&}
      \or\def\@tempa{}\fi\@tempa
{\hbox to 3cm{\phantom{\rm(\theequation)}
\draftnote \hfil}\hskip -3cm{\rm(\theequation)}}}

\def\eqnarray{\def\draftnote{{}}\global\@fewtabtrue
\stepcounter{equation}\let\@currentlabel=\theequation
\global\@eqnswtrue
\global\@eqcnt\z@\tabskip\@centering\let\\=\@eqncr
$$\halign to \displaywidth\bgroup\@eqnsel\hskip\@centering\@eqcnt\z@
  $\displaystyle\tabskip\z@{##}$&\global\@eqcnt\@ne
  \hskip 1\arraycolsep \hfil${##}$\hfil
  &\global\@eqcnt\tw@ \hskip 1\arraycolsep
$\displaystyle\tabskip\z@{##}$
\hfil  \tabskip\@centering&\global\@eqcnt\thr@@\llap{##}\tabskip\z@
\cr}

\def\endeqnarray{\@@eqncr\egroup
      \global\advance\c@equation\m@ne$$\global\@ignoretrue}

\def\@eqnnum{\hbox to 3cm{\phantom{\rm(\theequation)} \draftnote
                         \hfil}\hskip -3cm {\rm(\theequation)}}

\def\@@eqncr{\let\@tempa\relax
    \ifcase\@eqcnt \def\@tempa{& & &}\or \def\@tempa{& &}
      \or \def\@tempa{&}
      \or\def\@tempa{}
\fi\@tempa
\if@eqnsw
\if@fewtab\@eqnnum\fi
\stepcounter{equation}\fi\global
\@eqnswtrue\global\@eqcnt\z@\global\@fewtabtrue\cr}

\def\draftcite#1{\ifnum\draftcontrol=1#1\else{}\fi}

\def\@lbibitem[#1]#2{\item{}\hskip -3cm \hbox to 2cm
{\hfil$\scriptstyle\draftcite{#2}$}\hskip
1cm[\@biblabel{#1}]\if@filesw
     {\def\protect##1{\string ##1\space}\immediate
      \write\@auxout{\string\bibcite{#2}{#1}}}\fi\ignorespaces}

\def\@bibitem#1{\item\hskip -3cm \hbox to 2cm
{\hfil $\scriptstyle\draftcite{#1}$}\hskip 1cm
\if@filesw \immediate\write\@auxout
       {\string\bibcite{#1}{\the\value{\@listctr}}}\fi\ignorespaces}

\def\nsection#1{\section{#1}\setcounter{equation}{0}}

\font\tendl=msbm10  scaled \magstep1
\font\sevendl=msbm7 scaled \magstep1
\font\fivedl=msbm5 scaled \magstep1
\font\tengl=eufm10  scaled \magstep1
\font\sevengl=eufm7 scaled \magstep1
\font\fivegl=eufm5 scaled \magstep1

\newfam\dlfam \def\dl{\fam\dlfam\tendl} 
\textfont\dlfam=\tendl \scriptfont\dlfam=\sevendl
\scriptscriptfont\dlfam=\fivedl
\newfam\glfam  
\textfont\glfam=\tengl \scriptfont\glfam=\sevengl
\scriptscriptfont\glfam=\fivegl

\def\draftdate{\number\month/\number\day/\number\year\ \ \ \hourmin }

\global\def\draftcontrol{0}
\catcode`\@=12

\documentstyle[11pt]{article}

\def\theequation{{\thesection.\arabic{equation}}}

\newcommand{\be}{\begin{eqnarray}}
\newcommand{\en}{\end{eqnarray}\vs 0.5 cm}

\newcommand{\vs}{\vskip}
\newcommand{\hs}{\hspace}

\newcommand{\NC}{{{\dl C}}}
\newcommand{\qq}{\begin{eqnarray}}
\newcommand{\de}{\bar\partial}
\newcommand{\da}{\partial}
\newcommand{\ee}{{\rm e}}

\newcommand{\qqq}{\end{eqnarray}}

\newcommand{\CA}{{\cal A}}

\newcommand{\CC}{{\cal C}}

\newcommand{\CG}{{\cal G}}
\newcommand{\CH}{{\cal H}}
\newcommand{\CI}{{\cal I}}

\newcommand{\CO}{{\cal O}}

\newcommand{\s}{\hspace{0.05cm}}
\newcommand{\m}{\hspace{0.025cm}}

\newcommand{\bx}{{\bf x}}
\begin{document}
\title{\bf SU(2) WZNW model at higher genera from gauge
field functional integral}
\author{\ \\Krzysztof Gaw\c{e}dzki \\ C.N.R.S.,
I.H.E.S., Bures-sur-Yvette,
91440, France}
\date{ }
\maketitle

\vskip 1 cm

\begin{abstract}
We compute the gauge field functional integral giving the scalar
product of the \s$SU(2)\s$ Chern-Simons theory states on a Riemann
surface of genus \s$>\s1\s$. The result
allows to express the higher genera partition functions
of \m the \s$SU(2)\s$ WZNW conformal field theory by explicit
finite dimensional integrals. Our calculation may also
shed new light on the functional integral of the Liouville theory.
\end{abstract}

\vskip 0.9cm

\hs{2.8cm}{\it Dedicated to Ludwig Dmitrievich
Faddeev for the 60$^{th}$ birthday}
\vskip 1.7cm

\nsection{\hspace{-.6cm}.\ \ Introduction}
\vs 0.5cm

Since the 1967 seminal paper of Faddeev and Popov \cite{FP},
the functional integral has become the main tool in the
treatment of quantum gauge theories. The main breakthrough
which this paper has achieved was the realization
that the gauge invariance is not an obstruction but an aid
in the treatment of quantum gauge fields. Subsequently,
this idea has revealed its full power
and the gauge invariance has become
the cornerstone of modern theoretical physics.
\vs 0.5cm

In the present note, we shall discuss how the same idea
allows an explicit solution of the WZNW model
of conformal field theory on an arbitrary two-dimensional
surface.  To make things simpler, we shall limit the discussion
to the partition functions of the \s$SU(2)\s$ WZNW model
on closed compact Riemann surfaces \s$\Sigma\s$
of genus \s$>\s1\s$. The correlation functions at genus
zero were discussed along similar lines in
\cite{Quadr} for the \s$SU(2)\s$ case and in \cite{FalGaw0}
for the general compact groups. Higher genera, however, are
more difficult and took some time to understand.
The aim of this note is to present the main points of the
argument leaving the technical details to the forthcoming
publication \cite{Ja}. The complete work is a small {\it tour
de force}. This fact was hard to hide even in a softened
exposition which may only envy the work \cite{FP}
its striking simplicity.
\vs 0.5cm

Our approach is based on the relation between
the WZNW partition functions and the Schr\"{o}dinger
picture states of the Chern-Simons (CS) theory
\cite{WittenJones}\cite{Quadr}.
The partition function of the level \s$k\s$ (\s$k=1,2,\dots\s$)
\s$SU(2)\s$ WZNW model in an external
two-dimensional \s$su(2)\s$ gauge field
\s$A\equiv A_zdz+A_{\bar z}d\bar z\s$ is formally given
by the functional integral
\qq
Z(A)\s=\s\int\ee^{-k\m S(g,A)}\ Dg
\qqq
over \s$g:\Sigma\rightarrow SU(2)\s$. \s On the
other hand, the \s$SU(2)\s$ CS states are holomorphic
functionals \s$\Psi\s$ of the gauge field \s$A_{\bar z}\s$
s.\s t. for \s$h:\Sigma\rightarrow SL(2,\NC)\s$,
\qq
\ee^{-k\m S(h,A_{\bar z}d\bar z)}\ \Psi({}^{h^{-1}}
\hs{-0.17cm}A_{\bar z})\m\s=\s\Psi(A_{\bar z})\ ,
\label{CoV}
\qqq
where \s\s${}^{h^{-1}}
\hs{-0.17cm}A_{\bar z}\equiv h^{-1}\hs{-0.04cm}
A_{\bar z}\m h\m+\m h^{-1}\da_{\bar z}h\s$.
\m Above \s$S(\m\cdot\s,
\m\cdot\m)\s$ denotes the WZNW action in the external
gauge field \cite{WittenBos}. The CS states form a
finite-dimensional space with the dimension expressed by the
Verlinde formula \cite{Verl}. Their
scalar product is formally given by the functional
integral
\qq
\|\Psi\|^2\s\m=\s\int|\Psi(A_{\bar z})|^2\ \ee^{\s{k\over\pi}
\int_{_\Sigma}\hs{-0.03cm}{\rm tr}\s\s A_z A_{\bar z}\m d^2z}\ DA
\label{FI}
\qqq
(with the convention that \s$A_z=-A_{\bar z}^\dagger\s$ for
an \s$su(2)\s$ gauge field \s$A\s$)\m. \s The partition
function of the WZNW model is \cite{Quadr}
\qq
Z(A)\s\m=\s\sum\limits_{a,b}H^{ab}
\s\m\Psi_a(A_{\bar z})\s\s\overline{\Psi_b(A_{\bar z})}
\s\ \ee^{\s{k\over\pi}
\int_{_\Sigma}\hs{-0.03cm}{\rm tr}\s\s A_z A_{\bar z}\m d^2z}\ ,
\label{BF}
\qqq
where \s$(H^{ab})\s$ is the matrix inverse to that
of the scalar products \s$(\psi_a\m,\s\psi_b)\s$,
\s for an arbitrary basis of the CS states. Hence, in order
to find the WZNW partition functions, we should be able to
construct CS states and to compute their scalar product.
The calculation of the latter is another
exercise in the functional integration
over gauge fields and may be dealt with similarly as in
the original Faddeev-Popov's geometric argument.
The main point is to use the action \s$A_{\bar z}\mapsto
{}^{h^{-1}}\hs{-0.17cm}A_{\bar z}\s$ of the group
\s$\CG^\NC\s$ of the complex gauge transformations
in order to reduce the functional integral over
the space \s$\CA\s$ of gauge fields to the orbit space
\s$\CA/\CG^\NC\s$, which is, in fact, finite dimensional.
The main aspect differing our situation from that
considered by Faddeev and Popov is that neither
the integrated function nor the (formal) integration measure
\s$DA\s$ are invariant under
the complex gauge transformations. Nevertheless,
both transform in a controllable way. This difference
has to be properly accounted for.
\vs 0.5cm

The first step in the reduction of the functional
integral to the orbit space is to choose a slice
\s\s$\CA/\CG^\NC\ni n\s\smash{\mathop{\mapsto}\limits^s}
A(n)\in\CA\s\s$ which
cuts every orbit only once (generically) and to change
variables by parametrizing
\qq
A_{\bar z}\s=\s{}^{h^{-1}}\hs{-0.16cm}A_{\bar z}(n)\ .
\label{ChV}
\qqq
The Jacobian of the change of variables \s${\da\s(\s A\s )
\over\da\s(h,n)}\s$ plays then the role similar to that
of the ghost determinant in the approach of \cite{FP}
where the slice was fixed by constraining functions.
The next step is the calculation of the \s$h$-integral.
This step is not trivial, in contrast to the gauge
invariant situation where the gauge group integral
factors out as an overall constant. In fact, in
the case at hand, the integration over \s$h\s$,
more involved at genera \s$>\s1\s$ than for \s$g=0\s$
and \s$g=1\s$, leads to a result which may look
surprising at the first sight:
it gives not a function but a singular
distribution on the orbit space \s$\CA/\CG^\NC\s$.
Its treatment may shed some light on the
more complicated case of the Liouville theory
functional integral.
After the integration over \s$h\s$ is done, one is
left with an explicit finite-dimensional distributional
integral, essentially over the orbit space \s$\CA/\CG^\NC\s$,
\s which may be further reduced to the standard integral
over the support of the distribution.
\vs 0.5 cm

The paper is organized as follows. Section 2 is devoted
to the description of the slice \s$s:\CA/\CG^\NC\rightarrow
\CA\s$. \s The use of \s$s\s$ allows also a more
explicit characterization of the CS states. In Section 3,
we perform the change of variables (\ref{ChV}) and
study its Jacobian. In Section 4, we describe the calculation
of the \s$h$-integral over the \s$\CG^\NC$-orbits in \s$\CA\s$
which turns out to be iterative Gaussian. Finally,
in Section 5 we discuss the resulting finite-dimensional
integral representation for the CS scalar product.
\vskip 1cm

\nsection{\hspace{-.6cm}.\ \ The slice}
\vs 0.5cm

We would like to describe a surface \s$\{A(n)\}\s$ inside
\s$\CA\s$ which cuts each orbit of \s$\CG^\NC\s$
once (or a fixed number of times).
The orbit space \s$\CA/\CG^\NC\s$ (after
removal of a small subset of bad orbits) is an object well
studied in the mathematical literature
under the name of the moduli space
of (stable) holomorphic \s$SL(2,\NC)$-bundles
\cite{Sesh}\cite{NaraRama}. It has complex
dimension \s$3(g-1)\s$, \s where \s$g\s$ denotes the genus of
the underlying Riemann surface. This should then also be
the dimension of our slice of \s$\CA\s$. \s Let
\s$L_0\s$ be a spin bundle over \s$\Sigma\s$. \s$L_0\s$
is a holomorphic line bundle with local sections
\s$(dz)^{1/2}\s$. \s \s$L_0^{-1}\s$ will denote its
dual bundle. \s$L_0^{-1}\oplus L_0\s$ is a rank two
vector bundle over \s$\Sigma\s$ which, as a smooth bundle,
is isomorphic to the trivial bundle \s$\Sigma\times\NC^2\s$.
We shall fix a smooth isomorphism
\qq
U:\m L_0^{-1}\oplus L_0\rightarrow\Sigma\times\NC^2\ .
\qqq
We may assume that \s$U\s$ preserves the length
of the vectors calculated in \s$L_0^{-1}\oplus L_0\s$
with the help of a fixed Riemannian metric \s$\gamma=
\gamma_{z\bar z}\m dz\m d\bar z\s$ on \s$\Sigma\s$.
\s$U\s$ may be used to transport the gauge fields
\s$A_{\bar z}\s$ to the bundle \s$L_0^{-1}\oplus L_0\s$.
More exactly, the relation
\qq
B_{\bar z}\s=\s UA_{\bar z}U^{-1}\m+\m U\da_{\bar z}U^{-1}
\label{rel}
\qqq
establishes a one to one correspondence between
\s$A_{\bar z}d\bar z\s$ and
\qq
B_{\bar z}d\bar z\s=\left(\matrix{-a&b\cr c&a}\right)\ ,
\qqq
where \s$ a\s$ is a scalar 01-form on \s$\Sigma\s$
(\s$a\in\wedge^{01}\s$)\s,\ \s$b\s$ is an \s$L_0^{-2}$-valued
one (\s$b\in\wedge^{01}(L_0^{-2})\s)\s$ and \s$c\in
\wedge^{01}(L_0^2)\s$.
\vs 0.5cm

Let us present the surface \s$\Sigma\s$ as a polygone with sides
\s$a_1,\s b_1,\s a_1^{-1},\s b_1^{-1},\s\dots\m,\s a_g,\s b_g,\s
a_g^{-1},$
$b_g^{-1}\s$ given by the basic cycles. We shall fix
\s$x_0\in\Sigma\s$ in the corner where \s$b_g^{-1}\s$
and \s$a_1\s$ meet. Let
\s$\omega^i\s$,\ \s$i=1,\dots,g\s$, \s be the standard basis
of holomorphic forms with
\s$\smallint_{_{a_i}}\hs{-0.06cm}\omega^j=\delta^{ij}\s,$
\ \s$\smallint_{_{b_i}}\hs{-0.06cm}
\omega^j=\tau^{ij}\s$. \s We shall take
a slice of \s$\CA\s$ formed by the gauge fields
\s$A^{x,b}\s$ corresponding to
\qq
B_{\bar z}d\bar z\s=\left(\matrix{-a^x&b\cr 0&a^x}\right)
\equiv\s B^{x,b}_{\bar z}\ ,
\label{B}
\qqq
where
\qq
a^x\s=\s\pi\s(\hs{-0.03cm}\smallint_{_{x_0}}^{^x}
\hs{-0.06cm}\omega^i)\s
({_1\over^{{\rm Im}\s\tau}})_{_{ij}}\m\bar\omega^j\s\equiv
\s\pi\s(\hs{-0.03cm}\smallint_{_{x_0}}^{^x}\hs{-0.06cm}\omega)\s
({\rm Im}\s\tau)^{-1}\m\bar\omega\ .
\qqq
Let \s$L_x\s$ denote the holomorphic line
bundle obtained from \s$L_0\s$ by replacing its \s$\de\s$
operator by \s$\de+a^x\hs{-0.04cm}
\wedge\equiv\de_{L_x}\s$. We shall restrict
forms \s$b\s$ further by taking
one representative in each class
of \s\s${\wedge^{01}(L_0^{-2})\over(\de-2a^x\hs{-0.04cm}
\wedge\hs{-0.03cm})
\m(\CC^{\infty}(L^{-2}_0))}\m\cong\m H^1(L_x^{-2})\s$.
This may be done by imposing the condition
\qq
(\nabla+2\m\overline{a^x}\hs{-0.02cm}\wedge)\s b\s=\s0
\label{b}
\qqq
with \s$\nabla\s$ standing for the holomorphic covariant
derivative of the sections of \s$L_0^{-2}=T^{10}\Sigma\s$.
Finally, only one \s$b\s$ in each complex ray of solutions
of (\ref{b}) should be taken since
\qq
(\matrix{_\lambda&_0\cr^0&^{\lambda^{-1}}})\m\s B^{x,b}_{\bar z}
\m\m(\matrix{_{\lambda^{-1}}&_0\cr^0&^\lambda})\s=\s B^{x,\m
\lambda{\hs{-0.02cm}}^{^{_2}}\hs{-0.03cm}b}
\qqq
and, consequently, \s$A^{x,b}\s$ and \s$A^{x,\m\lambda b}\s$
are gauge related. It may be shown that the union
of the \s$\CG^\NC$-orbits passing through a slice of
\s$\CA\s$ constructed this way is dense
in \s$\CA\s$ and that the generic \s$\CG^\NC$-orbit
cuts the slice a fixed number of times \cite{Ja}.
Here, we shall content ourselves with the count of
dimensions. By the Riemann-Roch
theorem, the dimension of \s$H^1(L_x^{-2})\s$
is \s$3(g-1)\s$. \s The projectivization
subtracts one dimension which is restored by varying
\s$x\in\Sigma\s$.
\vs 0.5cm

The CS states \s$\Psi\s$ admit an explicit representation
when restricted to the slice described above. Let us put
\qq
\psi(x,b)\s=\s\ee^{-\pi\s(\hs{-0.02cm}\int_{_{x_0}}^{^x}
\hs{-0.1cm}\omega)\s({\rm Im}\s\tau)^{-1}\m(\hs{-0.02cm}
\int_{_{x_0}}^{^x}\hs{-0.1cm}\omega)}\ \m\ee^{\s{k\over\pi}
\int_{_\Sigma}\hs{-0.05cm}{\rm tr}\s A_{z}^0
A_{\bar z}^{x,b}\s d^2z}\ \Psi(A_{\bar z}^{x,b})\ ,
\qqq
where \s$A^0_z\equiv U\nabla_z U^{-1}\s$. \s$\psi\s$
is an analytic function of \s$x\s$ and \s$b\s$ depending
only of the class of \m\s$b\s$ modulo \s\s$(\de-2a^x\wedge)
(\CC^{\infty}(L_0^{-2}))\s$:
\qq
\hbox to 11.5cm{\hs{1.5cm}$\psi(x,\m b+
(\de-2a^x\hs{-0.04cm}\wedge)\m w)
\s=\s\psi(x,b)\ .$\hfill}\label{1}
\qqq
Under constant complex rescalings of \s$b\s$,
\qq
\hbox to 11.5cm{\hs{1.5cm}$\psi(x,\lambda b)
\s=\s\lambda^{k(g-1)}\s\psi(x,b)\ .$\hfill}\label{2}
\qqq
Under the action of the fundamental group
\s$\pi_1(\Sigma)\s$,
\qq
\hbox to 11.5cm{\hs{1.5cm}$\psi(a_jx,\m c_{a_j}^2 b)\s=\s
\nu(a_j)^k\s\s\psi(x,b)\ ,$\hfill}\label{3}\\
\hbox to 11.5cm{\hs{1.5cm}$\psi(b_jx,\m c_{b_j}^2 b)\s\s=\s
\nu(b_j)^k\s\s\ee^{-2\pi i k\tau^{jj}
\m-\m4\pi i k\hs{-0.04cm}\int_{_{x_0}}^{^x}\hs{-0.06cm}\omega^j}
\s\psi(x,b)\ ,$\hfill}\label{4}
\qqq
where, for each \s$p\in\pi_1(\Sigma)\s$, \s$c_p\s$
is a non-vanishing (univalued) function on \s$\Sigma\s$,
\qq
c_p(y)\s=\s\ee^{\m2\pi i\s\s{\rm Im}\s\s(\m(\hs{-0.02cm}\int_{_p}
\hs{-0.06cm}\omega)\s({\rm Im}\s\tau)^{-1}\m(\hs{-0.02cm}
\int_{_{x_0}}^{^y}\hs{-0.07cm}\bar \omega)\m)}
\qqq
and \s$\nu\s$ is a character of \s$\pi_1(\Sigma)\s$,
\qq
\nu(p)\s=\s\ee^{-{i\over 2\pi}\int_{_\Sigma}\hs{-0.04cm}R\s\ln\m c_p}
\ \prod\limits_{j=1}^g\s\m
W_{a_j}^{-{1\over\pi}\hs{-0.02cm}\int_{_{b_j}}
\hs{-0.1cm}c_p^{-1}dc_p}\hs{-0.02cm} W_{b_j}^{\m{1
\over\pi}\hs{-0.02cm}
\int_{_{a_j}}\hs{-0.1cm}c_p^{-1}dc_p}\ .
\qqq
In the last formula, the integral
\s$\int_{_\Sigma}\hs{-0.07cm}R\s\ln\m c_p\s$
is over the polygone representing the surface, \s$R\s$
is the metric curvature form of \s$\Sigma\s$ normalized
so that \s$\smallint_{_\Sigma}\hs{-0.04cm}R=4\pi i(g-1)\s$,
\s$\ln\s c_p(y)=\smallint_{_{x_0}}^{^y}\hs{-0.06cm}c_p^{-1}dc_p\s$
and \s$W_p\s$ stands for the holonomy of \s$L_0\s$
around the closed curve \s$p\s$.
One may identify functions \s$\psi\s$ satisfying relations (\ref{1}),
(\ref{2}), (\ref{3}) and (\ref{4}) with sections of the
\s$k^{\rm th}\s$ power of Quillen's determinant bundle \cite{Quill}
of the holomorphic family \s$(\s\de+B^{x,b}\s)\s$ of operators
in \s$L_0^{-1}\oplus L_0\s$. \s Holomorphic \s$\psi\s$'s \s
form a finite dimensional space and the relation
\s$\Psi\mapsto\psi\s$ realizes the space of CS states
as its, in general proper, subspace.
\vskip 1cm

\nsection{\hspace{-.6cm}.\ \ The change of variables}
\vs 0.5cm

We apply the change of variables (\ref{ChV})
with \s$A_{\bar z}(n)\equiv A_{\bar z}^{x,b}\s$
in the functional integral (\ref{FI}) giving the scalar
product of CS states. The Jacobian is
\qq
{\da\s(\s A\m\s )\over\da\s(h,n)}\s=\m\s{\rm det}\s(
D_{\bar z}(h,n)^\dagger\m
D_{\bar z}(h,n))\s\ \m\s{\rm det}\s(\m({_{\da A_{\bar z}}
\over{^{\da{n_\alpha}}}})^{^{\hs{-0.04cm}\perp}}\hs{-0.06cm},
\s({_{\da A_{\bar z}}
\over{^{\da{n_\beta}}}})^{^{\hs{-0.04cm}\perp}}\hs{-0.02cm})\ ,
\qqq
where \s\s$D_{\bar z}(h,n)\hs{-0.01cm}=\hs{-0.02cm}
h^{^{\hs{-0.02cm}{-1}}}\hs{-0.02cm}(\da_{\bar z}
+[A_{\bar z}(\hs{-0.02cm}n\hs{-0.02cm})\m,\s
\cdot\s\s])\m h\s\s\m$ and \m\s\s$({{\da A_{\bar z}}
\over{{\da{n_\alpha}}}}\hs{-0.03cm})^{^{\hs{-0.04cm}\perp}}\s$
denotes the component
of \s\m\s$h^{^{\hs{-0.03cm}{-1}}}\hs{-0.05cm}
{{\da A_{\bar z}(n)}\over{\da n_\alpha}}\m h\s$
perpendicular to the image of \s\s$D_{\bar z}(h,n)\s$.
The first determinant should be regularized (what specific
regularization is used does not matter as long as
it is insensitive to unitary conjugations of the
operator, like the zeta-function regularization).
The \s$h$-dependence of the regularized determinants
is given
by the global version of the non-abelian chiral anomaly formula
\cite{PolyWieg}
\qq
{\rm det}\s(
D_{\bar z}(h,n)^\dagger\m
D_{\bar z}(h,n))\s\ \s\m{\rm det}\s(\m({_{\da A_{\bar z}}
\over{^{\da{n_\alpha}}}})^{^{\hs{-0.04cm}\perp}}\hs{-0.06cm},
\s({_{\da A_{\bar z}}
\over{^{\da{n_\beta}}}})^{^{\hs{-0.04cm}\perp}}\hs{-0.02cm})
\hs{4cm} \cr
=\s\s\ee^{\s4\m S(hh^\dagger\hs{-0.02cm},\m A(n))}\ \s\s
{\rm det}\s(
D_{\bar z}(n)^\dagger\m
D_{\bar z}(n))\s\ \s\m{\rm det}\s(\m({_{\da A_{\bar z}(n)}
\over{^{\da{n_\alpha}}}})^{^{\hs{-0.04cm}\perp}}\hs{-0.03cm},
\s({_{\da A_{\bar z}(n)}
\over{^{\da{n_\beta}}}})^{^{\hs{-0.04cm}\perp}}\hs{-0.02cm})\ ,
\label{ChA}
\qqq
where \s\s$D_{\bar z}(n)\equiv\s D_{\bar z}(1,n)\s$.
The covariance property (\ref{CoV}) implies that
\qq
|\Psi(A_{\bar z})|^2\ \ee^{\s{k\over\pi}\hs{-0.03cm}
\int_{_\Sigma}\hs{-0.03cm}{\rm tr}\s\s A_z\hs{-0.02cm}
A_{\bar z}\s d^2z}\s=\s
|\Psi(A_{\bar z}(n)|^2\ \ee^{\s{k\over\pi}\hs{-0.03cm}
\int_{_\Sigma}\hs{-0.03cm}{\rm tr}\s\s A_z(n)\m
A_{\bar z}(n)\s d^2z}
\ \ee^{\s k\m S(hh^\dagger\hs{-0.02cm},\s A(n))}\ .
\qqq
Hence, after the change of variables,
the functional integral (\ref{FI}) becomes
\qq
\|\Psi\|^2\s\m=\s\int|\Psi(A_{\bar z}(n)|^2\ \ee^{\s{k\over\pi}
\hs{-0.03cm}\int_{_\Sigma}\hs{-0.03cm}{\rm tr}
\s\s A_z(n)\m A_{\bar z}(n)\s d^2z}
\ \ee^{\s(k+4)\m S(hh^\dagger\hs{-0.02cm},\s A(n))}\
\s{\rm det}\s(
D_{\bar z}(n)^\dagger\m
D_{\bar z}(n))\s\s\ \ \ \cr
\cdot\ \s{\rm det}\s(\m({_{\da A_{\bar z}(n)}
\over{^{\da{n_\alpha}}}})^{^{\hs{-0.04cm}\perp}}\hs{-0.03cm},
\s({_{\da A_{\bar z}(n)}
\over{^{\da{n_\beta}}}})^{^{\hs{-0.04cm}\perp}}\hs{-0.02cm})
\ \s D(hh^\dagger)\ \prod\limits_{\alpha}d^2n_\alpha\ ,\ \
\label{FI1}
\qqq
where we have used the \s$SU(2)\s$ gauge invariance of
the integral to factor out the integration over the \s$SU(2)$-valued
gauge transformations, like in Faddeev-Popov's case.
There remains, however, the integral over the field
\s$hh^\dagger\s$ effectively taking values in the hyperbolic
space \s$\CH\equiv SL(2,\NC)/SU(2)\s$. \s\s$D(hh^\dagger)\s$
should be viewed a formal product of the
\s$SL(2,\NC)$-invariant measures on \s$\CH\s$.
\vs 0.5cm

Working out the explicit form of various terms under
the integral in eq.\s\s(\ref{FI1}) for \s$A(n)\equiv A^{x,b}\s$
is a rather straightforward matter. One obtains
\qq
|\Psi(A_{\bar z}^{x,b})|^2\ \ee^{\s{k\over\pi}
\hs{-0.03cm}\int_{_\Sigma}\hs{-0.03cm}{\rm tr}
\s\s A_z^{x,b}\hs{-0.02cm} A_{\bar z}^{x,b}\s d^2z}
\s\s=\s\s\s
\ee^{\s{k\over \pi}\hs{-0.03cm}\int_{_\Sigma}
\hs{-0.05cm}{\rm tr}\s\m A^0_z\m(A^0_z)^\dagger\s d^2z}
\ \ee^{-4\pi \m k\s\m
(\int_{_{x_0}}^{x}\hs{-0.1cm}{\rm Im}\s\omega)\s\m
({\rm Im}\s\tau)^{-1}\m
(\int_{_{x_0}}^{x}\hs{-0.1cm}{\rm Im}\s\omega)}\ \cr
\cdot\ \ee^{\m{k\over 2\pi i}\hs{-0.03cm}\int_{_\Sigma}
\hs{-0.07cm}\langle b,\m\wedge b\rangle}\ \s|\psi(x,b)|^2
\ .\
\label{psi2}
\qqq
(Here and below, \s$\langle\m\cdot\s,\m\cdot\rangle\s$
denotes the hermitian structure induced by the
Riemannian metric of \s$\Sigma\s$  on the powers
of its canonical bundle.)
The determinant of the operator \s$D_{\bar z}(n)^\dagger\m
D_{\bar z}(n)\s$, unitarily equivalent to
the operator \s\s$(\da_{\bar z}+\m[B^{x,b}_{\bar z}
\m,\s\cdot\ ]\m)^\dagger
\m(\da_{\bar z}+\m[B^{x,b}_{\bar z}\m,\s\cdot\ ]\m)\s\s$
acting on smooth endomorphism of \s$L_0^{-1}\oplus L_0\s$,
\s may be found by performing the Gaussian integration
\qq
{\rm det}\m(\bar D_{\bar z}(n)^\dagger\m\bar D_{\bar z}(n))
\s\s=\s\int\ee^{-i\int_{_\Sigma}
(\m 2\m(\overline{\de X+Zb})\m\wedge
\m(\de X+Zb)\s
+\s\langle\m(\de-2a^x)Y-2Xb\m,\s\wedge
\s((\de-2a^x)Y-2Xb)\m\rangle\m)}
\cr
\cdot\ \ee^{\s\langle\m(\de+2a^x)Z\m,
\s\wedge\s(\de-2a^x)Z\m\rangle}
\m\ DY\ DX\ DZ\ \
\qqq
over the anticommuting ghost fields: the scalar \s$X\s$,
\s the \s$L_0^{-2}$-valued \s$Y\s$ and the
\s$L_0^2$-valued \s$Z\s$.
Formally, the computation may be done iteratively,
first over \s$Y\s$, then over \s$X\s$ and, at the end,
over \s$Z\s$. One obtains this way the product of
determinants
\qq
{\det}\s(\de_{L_x^{-2}}^\dagger\m\de_{L_x^{-2}})
\ \s{\rm det}'\m(-\Delta)\s\ {\det}'\m
(\de_{L_x^2}^\dagger\m\de_{L_x^2})\ ,
\qqq
where \s$\Delta\s$ is the scalar Laplacian on \s$\Sigma\s$,
\s decorated with zero mode terms. Some care should be taken
since the regularization (e.\s g. by the zeta-function
prescription) of the big determinant requires, besides
similar regularization of the product determinants,
also an additional term which may be found by demanding
that the result transforms as in (\ref{ChA}) under
the complex gauge transformations.
\vs 0.5cm

The final expression for the measure
\qq
d\mu(n)\s\s\s\equiv\s\s{\rm det}\s(
D_{\bar z}(n)^\dagger\m
D_{\bar z}(n))\s\s\ {\rm det}\s(\m({_{\da A_{\bar z}(n)}
\over{^{\da{n_\alpha}}}})^{^{\hs{-0.04cm}\perp}}\hs{-0.03cm},
\s({_{\da A_{\bar z}(n)}
\over{^{\da{n_\beta}}}})^{^{\hs{-0.04cm}\perp}}\hs{-0.02cm})
\ \prod\limits_{\alpha}d^2n_\alpha
\label{dmu}
\qqq
on the slice becomes rather simple. Fix \s$x\in\Sigma\s$.
Let \s\m$(b^\alpha)_{_{\alpha=1}}^{^{3(g-1)}}\s$
be an orthonormal basis (in the natural \s$L^2\s$ scalar product)
of \s$L_0^{-2}$-valued 01-forms \s$b\s$ solving  eq.\s\s(\ref{b}).
Similarly, \s let \s$(\kappa_r)_{_{r=1}}^{^{g-1}}\s$ be
an orthonormal basis of sections of \s$L^{-2}_0\s$
annihilated by \s$\nabla-2\overline{a^x}\hs{-0.02cm}\wedge\s$.
\s Set \s\s$M_{ir}^\alpha\m\equiv\smallint_{_\Sigma}\hs{-0.06cm}
\omega^i\langle\kappa_r,\m b^\alpha\rangle\s$.
Denote by \s$(z^\alpha)\s$ the complex coordinates
w.r.t. the basis \s$(b^\alpha)\s$ on the space
of \s$b\s$ satisfying eq.\s\s(\ref{b}). One may use
\s$x\in\Sigma\s$ and the homogeneous coordinates \s$z^\alpha\s$
to parametrize the slice of \s$\CA\s$ and
\qq
d\mu(x,b)\s\s=\s\s{\rm const.}\ \s{_{{\rm area}(\Sigma)}
\over^{{\rm det}\s({\rm Im}\s\tau)}}\ \s
{\det}\s(\de_{L_x^{-2}}^\dagger\m\de_{L_x^{-2}})
\ \s{\rm det}'\m(-\Delta)\s\ {\det}'\m
(\de_{L_x^2}^\dagger\m\de_{L_x^2})\
\s\ee^{\m{2\over\pi i}\int_{_\Sigma}\hs{-0.04cm}
\langle b,\m\wedge b\rangle}\ \ \cr
\cdot\ |\s\epsilon_{\alpha_1,\dots,\alpha_{3(\chi-1)}}
\s\m z^{\alpha_1}\m dz^{\alpha_2}\hs{-0.03cm}\wedge\m
\dots\m\wedge dz^{\alpha_{3(\chi-1)}}\s|^2\ \s
|\sum\limits_{j=1}^g\s{\rm det}\s(\m
\sum_\alpha M_{ir}^\alpha
z^\alpha\m)_{\hs{-0.01cm}_{i\not=j}}
\ \omega^j(x)\s|^2\ .\
\label{dmu1}
\qqq
Notice that \s$d\mu(n)\s$ contains as a factor
the natural measure
on the projectivization of the space of solutions of
eq.\s\s(\ref{b}).
\vskip 1cm

\nsection{\hspace{-.6cm}.\ \ Integral over the gauge orbits}
\vs 0.5cm

It remains to compute the functional integral
\qq
Z_{_{\CH}}(A^{x,\m b})\s\s\equiv\s\int\ee^{\m(k+4)\m
S(hh^\dagger,\m A^{x,b})}\ D(hh^\dagger)
\label{HFI}
\qqq
giving the genus \s$g\s$ partition function of a
(nonunitary) WZNW model
with fields taking values in the hyperbolic space \s$\CH\s$,
first considered in \cite{GK}.
The calculation of the functional integral (\ref{HFI})
is the crucial step of the argument. The fields \s$hh^\dagger\s$
may be uniquely parametrized by real functions \s$\varphi\s$
and sections \s$w\s$ of \s$L_0^{-2}=T^{10}\Sigma\s$ by
writing
\qq
hh^\dagger\s\s=\s\s U\left(\matrix{1&{\ee^\varphi w}\cr0&1}\right)
\left(\matrix{{\ee^\varphi}&0\cr0&{\ee^{-\varphi}}}\right)
\left(\matrix{1&0\cr{\ee^\varphi w^\dagger}&1}\right) U^{-1}
\label{hh}
\qqq
where \s$w^\dagger\s$ is the section of \s$L_0^2\s$
obtained by contracting the vector field \s$\bar w\s$
with the Riemannian metric. Rewritten in terms of \s$\varphi\s$
and \s$w\s$, the functional integral (\ref{HFI}) becomes
\qq
Z_{_{\CH}}(A^{x,\m b})&=&\ee^{-{k+4\over 2\pi i}\hs{-0.1cm}
\int_{_\Sigma}\hs{-0.1cm}
\langle b\s,\s\wedge\m b\rangle}\ \ \hs{6.5cm}\cr
&\cdot&\int\ee^{{k+4\over 2\pi i}\int_{_\Sigma}\hs{-0.02cm}[\s
-\varphi\s(\da\de\varphi+R\m)\s\m+
\m\s\langle\s\ee^{-\varphi}b
+(\de+(\de\varphi))\m w\s,\m\s\wedge\m(
\ee^{-\varphi}b+
(\de+(\de\varphi))\m w)\s\rangle\m]}\ Dw\s\s D\varphi\s ,\ \ \ \
\label{HFIE}
\qqq
where \s$\de\equiv\de_{L_x^2}\s$ when acting on \s$w\s$.
The \s$w$-integral is Gaussian and may be easily performed:
\qq
\CI_w&\equiv&\int\ee^{{k+4\over 2\pi i}\int_{_\Sigma}\hs{-0.02cm}
\langle\s\ee^{-\varphi}b
+(\de+(\de\varphi))\m w\s,\m\s\wedge\m(
\ee^{-\varphi}b+
(\de+\de(\varphi))\m w)\s\rangle}\ Dw\hs{2cm}\cr
&=&\ee^{{k+4\over 2\pi i}\int_{_\Sigma}\hs{-0.02cm}
\langle\s P_\varphi(\ee^{-\varphi}b)\s,
\s\m\wedge\m P_\varphi(\ee^{-\varphi}b)
\s\rangle}\ \s{\rm det}\left((\de_{L_x^2}+(\de\varphi))^\dagger\s
(\de_{L_x^2}+(\de\varphi))\right)^{-1}\ ,
\label{HH0}
\qqq
where \s$P_{\varphi}\s$ denotes the orthogonal projection
on the kernel of \s\s$(\de_{L_x^2}+(\de\varphi))^\dagger\s$.
Since the latter is spanned by the vectors \s$\ee^\varphi
b^\alpha\s$,
\qq
i\smallint_{_\Sigma}\hs{-0.02cm}
\langle\s P_\varphi(\ee^{-\varphi}b)\s,
\s\m\wedge\m P_\varphi(\ee^{-\varphi}b)
\s\rangle\s\equiv\s\|P_\varphi(\ee^{-\varphi} b)\|^2\ \cr
=\s
\overline{(b^\alpha\m,\s b)}\s\m(H_\varphi^{-1})_{\alpha\beta}\m
\s(b^\beta\m,\s b)\s=\s\bar z^\alpha\s\m
(H_\varphi^{-1})_{\alpha\beta}\m
\s z^\beta\ ,
\qqq
where \s$H_\varphi\s$ is the matrix of scalar products
\s\s$(\ee^{\varphi}b^\alpha\m,\s\ee^\varphi b^\beta)\s$.
The factor \s$\ee^{-{k+4\over 2\pi}
\|P_\varphi(\ee^{-\varphi}b)\|^2}\s\s$ is the classical
value of the \s$w$-integral.
One may rewrite it as a finite-dimensional integral
\qq
\ee^{-{k+4\over 2\pi}
\|P_\varphi(\ee^{-\varphi}b)\|^2}\s
=\s{\rm const}.\ \s{\rm det}\s(\m H_\varphi)\s
\int\ee^{-{2\pi\over k+4}\s\m\bar c_\alpha\s
(H_\varphi)^{\alpha\beta}\hs{0.015cm}
\s c_\beta\m\s+\s\m i\s \bar c_\alpha z^\alpha\s\m
+\s\s i\s c_\alpha\bar z^\alpha}
\ \prod\limits_\alpha d^2c_\alpha\ .
\label{HH1}
\qqq
By the global version of the abelian chiral anomaly
formula,
\qq
{\rm det}\s(\m H_\varphi)\ \s{\rm det}
\left((\de_{L^{-2}_x}+(\de\varphi))^\dagger\s
(\de_{L^{-2}_x}+(\de\varphi))\right)^{-1}\hs{2.2cm}\hs{2.5cm}\cr
=\s\s\ee^{-{1\over\pi i}\int_{_\Sigma}\hs{-0.04cm}
\da\varphi\wedge\de\varphi\s\s+
\s\s{3\over 2\pi i}\int_{_\Sigma}\hs{-0.05cm}\varphi\m R}
\ \s
{\rm det}\s(\m\de_{L^{-2}_x}^\dagger\s
\de_{L^{-2}_x}\m)^{-1}
\label{HH2}
\qqq
(recall that \s$H_0\s$ is the unit matrix).
Gathering eq.\s\s(\ref{HH0}), (\ref{HH1}) and (\ref{HH2}),
one obtains
\qq
\CI_w\s=\s{\rm const}.\s\ {\rm det}\s(\m\de_{L^{-2}_x}^\dagger\s
\de_{L^{-2}_x}\m)^{-1}
\ \s\s\ee^{-{1\over\pi i}\int_{_\Sigma}\hs{-0.04cm}
\da\varphi\wedge\de\varphi\s\s+
\s\s{3\over 2\pi i}\int_{_\Sigma}\hs{-0.05cm}\varphi\m R}\ \ \cr
\cdot\s\int\ee^{-{2\pi\over k+4}\s\m\bar c_\alpha\s
(H_\varphi)^{\alpha\beta}
\m\s c_\beta\m\s+\s\m i\s \bar c_\alpha z^\alpha\s\m
+\s\s i\s c_\alpha\bar z^\alpha}
\ \prod\limits_\alpha d^2c_\alpha\ .
\qqq
Note that the right hand side which, together with
the \s$\varphi$-terms left over in (\ref{HFIE}), has to be integrated
over \s$\varphi\s$ contains a Liouville-type term
\qq
\ee^{-{2\pi\over k+4}\s\m\bar c_\alpha\s
(H_\varphi)^{\alpha\beta}
\m\s c_\beta}\s=\s\ee^{-{2\pi i\over k+4}\int_{_\Sigma}
\hs{-0.06cm}\ee^{2\varphi}\s\langle c_\alpha b^\alpha,\s
c_\beta b^\beta\rangle}
\qqq
notorious for causing problems in the functional integration.
Indeed, with the \s$w$-integral done, the \s$\varphi$-integral
takes the form
\qq
\int\ee^{\m{k+2\over 2\pi i}\int_{_\Sigma}\hs{-0.04cm}
\da\varphi\wedge\de\varphi\s\s-{k+1\over 2\pi i}
\int_{_\Sigma}\hs{-0.06cm}\varphi\s R\s\s-
\s\s{2\pi\over k+4}\s\m\bar c_\alpha\s
(H_\varphi)^{\alpha\beta}
\m\s c_\beta}\ D\varphi\s\s\ \equiv\s\ \s\CI_\varphi
\qqq
which, unlike at low genera, is of the Liouville theory
type, not a Gaussian one.
A possible approach to such an integral is to
try to get rid of the Liouville term by integrating out
the zero mode \s\s$\varphi_0\equiv({\rm area}
\m(\Sigma))^{-{1\over2}}
\smallint_{_\Sigma}\hs{-0.06cm}\varphi\s da\s\s$ of \s$\varphi\s$
(\s$da\s$ denotes the metric volume measure on \s$\Sigma\s$)\s.
This was tried in the Liouville theory in \cite{GTW}
and, supplemented with rather poorly understood formal tricks,
has led in \cite{GL} to the functional integral
calculation of three-point functions for the minimal models
coupled to gravity. Multiplying \s$\CI_\varphi\s$
by \s\s$1\m=\m
({\rm area})^{1\over2}\smallint\delta
(\varphi_0-u\s({\rm area})^{1\over2})\s du\s$, \s\s changing
the order of integration, shifting \s$\varphi_0\s$ to
\s$\varphi_0+u\s$ and setting \s$M\equiv(k+1)(g-1)\s$,
one obtains
\qq
\CI_{\varphi}\s\s=\s\s({\rm area})^{1/2}
\s\int\hs{-0.03cm}\ee^{-2uM}\ \hs{-0.04cm}
\ee^{\m{k+2\over 2\pi i}\hs{-0.05cm}\int_{_\Sigma}\hs{-0.1cm}
\da\varphi\wedge\de\varphi\s
-\s{k+1\over 2\pi i}\int_{_\Sigma}
\hs{-0.08cm}\varphi\m R\s\m-\s\m{2\pi\over k+4}
\s\m\ee^{2u}\s\bar c_\alpha\s(H_\varphi)^{\alpha\beta}\m c_\beta}
\s\m du\ \delta(\varphi_0)\ D\varphi\hs{-0.01cm}\hs{0.8cm}\cr
=\s\s{_1\over^2}\s({\rm area})^{1/2}\s\Gamma(-M)\s\s
({_{2\pi}\over^{k+4}})^{M}\hs{-0.08cm}\int\hs{-0.05cm}
\ee^{\m{k+2\over 2\pi i}
\s\int_{_\Sigma}\hs{-0.1cm}
\da\varphi\wedge\de\varphi\s
-\m{k+1\over 2\pi i}\int_{_\Sigma}
\hs{-0.08cm}\varphi\m R}
\s (\m\bar c_\alpha\s(H_\varphi)^{\alpha\beta}
c_\beta\m)^{M}
\s\s\delta(\varphi_0)\s\s D\varphi\s.\hs{0.5cm}\
\label{GL}
\qqq
Hence the integration over the zero mode of \s$\varphi\s$
diverges but may be easily (multiplicatively) renormalized
by removing the overall divergent factor \s$\Gamma(-M)\s$.
Now, the \s$c$-integral is easy to perform\m:
\qq
\int(\m\bar c_\alpha\s(H_\varphi\m)_{\alpha\beta}\s c_\beta)^{M}
\s\s\ee^{\s i\s \bar c_\alpha z^\alpha\m+\s\m i\s c_\alpha
\bar z^\alpha}\m\prod\limits_\alpha
d^2c_\alpha\hs{5cm}\cr
=\ (2\pi)^{6(g-1)}\s\m
(-(H_\varphi)^{\alpha\beta}\m\da_{z^\alpha}
\da_{\bar z^\beta}\hs{-0.03cm})^{M}\s\prod\limits_\alpha\hs{-0.05cm}
\delta(z^\alpha)\s.
\qqq
Gathering the above results, we obtain the following
``Coulomb gas representation'' for the higher genus
partition function of the \s$\CH$-valued
WZNW model\m:

\vs 3cm

\qq
\int\hs{-0.05cm}\ee^{\m(k+4)\m S(hh^\dagger\hs{-0.02cm},
\m A^{x,b})}\ D(hh^\dagger)
\s=\s{\rm const}.\ \s({\rm area})^{1/2}\
{\rm det}\s(\m\de_{L^{-2}_x}^\dagger\m
\de_{L^{-2}_x})^{-1}\s\s
\ee^{-{k+4\over 2\pi i}\hs{-0.05cm}\int_{_\Sigma}\hs{-0.05cm}
\langle b\s,\s\wedge\m b\rangle}\ \ \s\ \cr
\cdot\s\left(
\int\ee^{\m{k+2\over 2\pi i}\int_{_\Sigma}\hs{-0.1cm}\s
\da\varphi\wedge\de\varphi\s\s
-\s\s{k+1\over 2\pi i}\int_{_\Sigma}
\hs{-0.05cm}\varphi\m R}\ (\hs{-0.04cm}-(H_\varphi)^{\alpha\beta}
\m\da_{z^\alpha}
\da_{\bar z^\beta}\hs{-0.03cm})^{M}\ \delta(\varphi_0)\s\
D\varphi\right)
\prod\limits_\alpha\hs{-0.03cm}\delta(z^\alpha)\ \ \s\m\ \cr\cr
=\s{\rm const}.\ ({\rm area})^{1/2}\
{\rm det}\s(\m\de_{L^{-2}_x}^\dagger\m
\de_{L^{-2}_x}\hs{-0.04cm})^{-1}\s\s
\ee^{-{k+4\over 2\pi i}\hs{-0.05cm}\int_{_\Sigma}\hs{-0.05cm}
\langle b\s,\s\wedge\m b\rangle}\m\bigg(\hs{-0.09cm}
\prod\limits_m
{_1\over^i}\s\da_{z^{\alpha_m}}\da_{\bar z^{\beta_m}}
\hs{-0.09cm}\prod\limits_\alpha
\delta(z^\alpha)\hs{-0.03cm}\bigg)\ \ \ \cr
\cdot\s \int\left(
\int\ee^{\m{k+2\over 2\pi i}\int_{_\Sigma}\hs{-0.1cm}\s
\da\varphi\wedge\de\varphi\s\s
-\s\s{k+1\over 2\pi i}\int_{_\Sigma}
\hs{-0.05cm}\varphi\m R\s\s+\s\s2\sum_{_m}\hs{-0.1cm}
\varphi(x_m)}\ \delta(\varphi_0)\ D\varphi\right)
\prod\limits_m\langle b^{\alpha_m},\s b^{\beta_m}
\hs{-0.05cm}\rangle(x_m\hs{-0.04cm})\ ,\
\label{PFN}
\qqq
where \s$m\s$ runs from \s$1\s$ to \s$M\s$.
The \s$\varphi\s$ integral is now purely Gaussian
and easily doable, provided that we Wick order
the ``screening charge'' insertions
\s$\ee^{2\varphi(x_m)}\s$ (this is, again,
a multiplicative renormalization). In the end, we obtain
\qq
Z_{_{\CH}}(A^{x,\m b})\s\s\s=\s\s\s{\rm const}.\hs{9.3cm}\cr\cr
\cdot\ \s{\rm det}\s(\m\de_{L^{-2}_x}^\dagger\m
\de_{L^{-2}_x})^{-1}\s
\left({_{{\rm det}'\m(-\Delta)}\over^{{\rm area}\m(\Sigma)}}
\right)^{\hs{-0.09cm}-1/2}\s
\ee^{-{k+4\over 2\pi i}\hs{-0.05cm}\int_{_\Sigma}\hs{-0.05cm}
\langle b\s,\s\wedge\m b\rangle}\s
\s\bigg(\hs{-0.09cm}
\prod\limits_m
{_1\over^i}\s\da_{z^{\alpha_m}}\da_{\bar z^{\beta_m}}
\hs{-0.09cm}\prod\limits_\alpha
\delta(z^\alpha)\hs{-0.03cm}\bigg)\s\ \ \cr
\cdot\s\int\bigg(\prod\limits_{m_1\not=m_2}
\hs{-0.27cm}\ee^{-{4\pi\over k+2}\m G(x_{m_1},\m x_{m_2})}\bigg)\s
\bigg(\prod\limits_m \ee^{-{4\pi\over k+2}
\s:\m G(x_m,\m x_m)\m:}
\s\m \langle b^{\alpha_m},\s b^{\beta_m}
\hs{-0.05cm}\rangle(x_\alpha\hs{-0.04cm})\bigg)\ ,
\label{HYPF}
\qqq
where \s$G(\m\cdot\s,\m\cdot\m)\s$ is the Green
function of the scalar Laplacian \s$\Delta\s$ on \s$\Sigma\s$
chosen so that \s\s$\smallint_{_\Sigma}\hs{-0.08cm}
G(\m\cdot\s,y)\s
R(y)=0\s$.  \s\s$:\hs{-0.02cm}G(y,y)\hs{-0.02cm}:
\s\s\equiv\s\lim\limits_{\epsilon\rightarrow 0}
\s\m(\m G(y,y')-{1\over 2\pi}\s\ln\s\epsilon\s)\s\s$ where
\s$\epsilon=d(y,y')\s$ is the distance between \s$y\s$
and \s$y'\s$.
\vs 0.5cm

Formula (\ref{HYPF}) reduces the functional integral
over \s$hh^\dagger\s$ to a finite dimensional integral
over positions \s$x_m\hs{-0.08cm}\in\hs{-0.06cm}
\Sigma\m\s$ of \s$M\s$ screening
charges. The integrand is a smooth function except for
\s$\CO(d(x_{m_1},\m x_{m_2})^{-{4\over k+2}})\s$ singularities
at coinciding points. Power counting shows, that the
integral converges for \s$g=2\s$ but for higher
genera it diverge unless special combinations
of forms \m$\langle b^{\alpha_m},\s b^{\beta_m}
\hs{-0.05cm}\rangle(x_m\hs{-0.04cm})\s$
are integrated.
Another feature of the right hand side of eq.\s\s(\ref{HYPF})
is even more surprising in the candidate for a
partition function:
its dependence on the external field \s$A^{\bx,b}\s$
is not functional but distributional\m! The entire
dependence on \s$b\s$ resides in the term \s\s$\prod\limits_\alpha
\da_{z^{\alpha_m}}\da_{\bar z^{\beta_m}}
\hs{-0.09cm}\prod\limits_\alpha
\delta(z^\alpha)\s\s$ (recall that \s$z^\alpha\equiv
(b^\alpha\m,\s b)\s$)\s.
\s\s This fact is
not so astonishing since the
partition function of the \s$\CH$-valued WZNW
model may be expected, by formal arguments
similar to the ones used in \cite{Wittenfact},
to be the hermitian square of a
holomorphic section of a negative power of the
determinant bundle. But there are no such global sections
but only distributional solutions
of the corresponding Ward identities. The right hand
side of (\ref{HYPF}) is one of them.
\vskip 1cm

\nsection{\hspace{-.6cm}.\ \ The scalar product formula}
\vs 0.5cm

In view of the results (\ref{dmu1}) giving the integration
measure on the slice and (\ref{HYPF}) computing
the integral along the \s$\CG^\NC$-orbits, the functional
integral expression (\ref{FI1}) for the scalar product
of the CS states reduces to
\qq
\|\Psi\|^2\ \s\s=\ \s\s{\rm const}.\s\ {\rm det}\s
({\rm Im}\s\tau)^{-1}\s\s\m\left({_{{\rm det}'\m
(-\Delta)}\over^{{\rm area}\m(\Sigma)}}
\right)^{\hs{-0.07cm}1/2}\ee^{\s{k\over \pi}\hs{-0.03cm}
\int_{_\Sigma}\hs{-0.05cm}{\rm tr}\s\m
A^0_z\m(A^0_z)^\dagger\s d^2z}\hs{3cm}\cr\cr
\cdot\s\int\ \ee^{-4\pi \m k\s\m
(\int_{_{x_0}}^{x}\hs{-0.1cm}{\rm Im}\s\omega)\s\m
({\rm Im}\s\tau)^{-1}\m
(\int_{_{x_0}}^{x}\hs{-0.1cm}{\rm Im}\s\omega)}
\ \s|\psi(x,\m b)|^2
\ |\sum\limits_{j=1}^g\s{\rm det}\m(
\sum_\alpha\hs{-0.05cm}M_{ir}^\alpha
z^\alpha\m)_{\hs{-0.01cm}_{i\not=j}}
\ \omega^j(x)\s|^2\ \hs{0.5cm}\ \s\s\cr
\cdot\ \s{\rm det}'\m(\m\de_{L^{2}_x}^\dagger\m
\de_{L^{2}_x})\s\bigg(\hs{-0.09cm}\prod\limits_m
{_1\over^i}\s\da_{z^{\alpha_m}}\da_{\bar z^{\beta_m}}
\hs{-0.09cm}\prod\limits_\alpha
\delta(z^\alpha)\hs{-0.03cm}\bigg)\s
\s|\s\epsilon_{\alpha_1,\dots,\m\alpha_{3(g-1)}}
\s z^{\alpha_1} dz^{\alpha_2}\hs{-0.06cm}\wedge
\dots\wedge\hs{-0.03cm} dz^{\alpha_{3(g-1)}}\s|^2\s\s\ \s\cr
\cdot\s\bigg(\hs{-0.08cm}\prod\limits_{m_1\not=m_2}
\hs{-0.3cm}\ee^{-{4\pi\over k+2}\s G(x_{m_1},\m x_{m_2})}\bigg)\m
\bigg(\prod\limits_m \ee^{-{4\pi\over k+2}\s:
\m G(x_m,\m x_m)\m:}\m\langle b^{\alpha_m},\s b^{\beta_m}
\hs{-0.05cm}\rangle(x_m\hs{-0.04cm})\bigg)\s.\ \
\qqq
The integral is, for fixed \s$x\in\Sigma\s$, over
the \s$(3g-4)$-dimensional projective space
with homogeneous coordinates \s$(z^\alpha)\s$ and over the
Cartesian product of \s$M\equiv(k+1)(g-1)\s$ copies of
\s$\Sigma\s$ (positions \s$x_m\s$ of the screening charges).
Finally, one should integrate over \s$x\in\Sigma\s$.
\vskip 0.5cm

The \s$(z^\alpha)$-integral should be interpreted
as
\qq
\int\limits_{\NC^{3(g-1)}}
\hs{-0.3cm}|P(z)|^2\s
\s\bigg(\hs{-0.09cm}\prod\limits_m
{_1\over^i}\s\da_{z^{\alpha_m}}\da_{\bar z^{\beta_m}}
\hs{-0.09cm}\prod\limits_\alpha
\delta(z^\alpha)\hs{-0.03cm}\bigg)\s
\s d^{\m6(g-1)}z\s\s=\s\s\prod\limits_m({_1\over^i}
\s\da_{z^{\alpha_m}}
\da_{\bar z^{\beta_m}})\Big|_{_{z=0}}\s|P(z)|^2
\label{delt}
\qqq
where \s\s$P(z)\m=\m\psi(x,\m b)\s
\sum\limits_{j=1}^g\s{\rm det}\m(
\sum_{_\alpha}\hs{-0.1cm}M_{ir}^\alpha
z^\alpha\m)_{\hs{-0.01cm}_{i\not=j}}\s\m\omega^j(x)\s\s$ is
a homogeneous polynomial in \s$(z^\alpha)\s$ of degree
\s$M\s$. \s Formally, the latter integral differs from the
one involving the volume form on \s$P\NC^{3g-4}\s$
by an infinite constant, which may be interpreted as
the factor \s$\Gamma(-M)\s$ which appeared in
the integral over the zero mode of the field \s$\varphi\s$.
With the \s$z$-integration given by eq.\s\s(\ref{delt}),
one obtains the following formula
for the scalar product of CS states:
\qq
\|\Psi\|^2\ \s=\ \s{\rm const}.\s\ {\rm det}\s
({\rm Im}\s\tau)^{-1}\s\s\left({_{{\rm det}'\m
(-\Delta)}\over^{{\rm area}\m(\Sigma)}}
\right)^{\hs{-0.07cm}1/2}\ee^{\s{k\over \pi}\hs{-0.03cm}
\int_{_\Sigma}\hs{-0.05cm}{\rm tr}\s\m
A^0_z\m(A^0_z)^\dagger\s d^2z}\hs{1.18cm}\s\cr
\cdot\s\int\s\prod\limits_m({_1\over^i}\s\da_{z^{\alpha_m}}
\da_{\bar z^{\beta_m}})\Big|_{_{z=0}}\s\bigg(
|\psi(x,\m b)|^2
\ |\sum\limits_{j=1}^g\s{\rm det}\m(
\sum_{_\alpha}M_{ir}^\alpha
z^\alpha\m)_{\hs{-0.01cm}_{i\not=j}}
\ \omega^j(x)\s|^2\bigg)\ \s\s\m\cr
\cdot\s\bigg(\hs{-0.08cm}\prod\limits_{m_1\not=m_2}
\hs{-0.3cm}\ee^{-{4\pi\over k+2}\s G(x_{m_1},\m x_{m_2})}\bigg)\m
\bigg(\prod\limits_m \ee^{-{4\pi\over k+2}\s:
\m G(x_m,\m x_m)\m:}\m\langle b^{\alpha_m},\s b^{\beta_m}
\hs{-0.05cm}\rangle(x_m\hs{-0.04cm})\bigg)\ \s\s\m\cr
\cdot\ \ee^{-4\pi \m k\s\m
(\int_{_{x_0}}^{x}\hs{-0.1cm}{\rm Im}\s\omega)\s\m
({\rm Im}\s\tau)^{-1}\m
(\int_{_{x_0}}^{x}\hs{-0.1cm}{\rm Im}\s\omega)}
\ \s{\rm det}'\m(\m\de_{L^{2}_x}^\dagger\m
\de_{L^{2}_x})\ ,\s
\label{final}
\qqq
where the numerical constant depends on the genus \s$g\s$ and
on the level \s$k\s$. \s The integration in (\ref{final}) is over
\s$x_m\s$, \s$m=1,\dots,(k+1)(g-1)\s$, \s and
over \s$x\s$, \s all in \s$\Sigma\s$. It is not difficult to
check that upon the Weyl rescalings \s$\gamma\mapsto\ee^{\sigma}
\gamma\s$ of the Riemannian metric
on \s$\Sigma\s$, the right hand side of (\ref{final}) (with
the zeta-function regularized determinants)
changes by the factor \s\s$\ee^{\m{k\over 8\pi i(k+2)}\int_{_\Sigma}
\hs{-0.07cm}({1\over 2}\m\da\sigma\wedge\de\sigma+\sigma R)}\s$.
\s This produces the correct value of the Virasoro central
charge of the WZNW partition functions given by eq.\s\s(\ref{BF}).
\vs 0.5cm

The arguments which led to eq.\s\s(\ref{final}) were clearly formal
and the treatment of the Liouville integral might have
looked particularly suspicious. Fortunately, one may
do the calculation in a different, more satisfying manner.
If the \s$z$-integral over \s$P\NC^{3g-4}\s$ is done
just after the \s$w$-integration and before
the one over \s$\varphi\s$ then
the final result is exactly as above but no infinite
constants (apart from the Wick ordering
ones) appear in the intermediate
steps. This way, it is rather the convergent integration
over the (part of) the modular degrees of freedom,
not the divergent \s$\varphi_0\s$ integral, which removes the
cumbersome Liouville-type terms from the effective
action for \s$\varphi\s$. This is an important lesson
to learn from the above calculation. It is plausible that
similar arguments may be used to substantiate the Goulian-Li
trick \cite{GL} in the gravity case.
\vskip 0.5cm

Similarly as for the genus zero case discussed in \cite{Quadr},
the natural conjecture is that the integral on the right hand
side converges if and only if the function \s$\psi\s$
defines a global non-singular CS state \s$\Psi\s$.
It is clear from the form (\ref{final}) of the scalar
product that finiteness of the screening
charge integral in (\ref{final}) imposes,
in general, conditions for the Taylor coefficients
of \s$\psi(x,\m b)\s$ at \s$b=0\s$ if \s$g>2\s$.
We shall postpone the study of these ``fusion rule
conditions'' to a future work. The case \s$g=2\s$ is
specially accessible since there
exists a simple global picture of the moduli
space of \s$SL(2,\NC)$-bundles
(it is the projectivization of the 4-dimensional
space of degree 2 theta-functions)\footnote{I thank
O. Garcia-Prada for explaining this work to me} \cite{NaraRama}
and of the space of CS states (homogeneous polynomials
of order \s$k\s$ on the same space).
\vs 0.3cm

If our conjecture is true, then the formula (\ref{final})
defines a hermitian structure on the holomorphic vector
bundle with the fibers given by the spaces of the CS states
and the base by the moduli space of complex curves.
Such a hermitian structure induces a holomorphic
hermitian connection. This connection should coincide
with the generalization to higher genera
of the Knizhnik-Zamolodchikov
connection studied in \cite{Denis}\cite{Kar}\cite{AxWitt}.
The latter has been constructed in \cite{Hitch}
in the geometric terms, and it is challenging to find
an interpretation for eq.\s\s(\ref{final})
in terms of the moduli space geometry.
\vs 2cm

\end{document}